\documentclass[12pt,oneside]{revtex4}
\usepackage{amssymb}
\usepackage{footmisc}
\usepackage{natbib}
\usepackage{anysize}
\usepackage{graphicx}
\usepackage{amsmath}
\usepackage{wasysym}

\usepackage{setspace}
\usepackage[bookmarks = false]{hyperref}
\usepackage{color}
\usepackage{subfigure}
\usepackage{multirow}
\usepackage[T1]{fontenc}
\renewcommand{\baselinestretch}{1.5}
\renewcommand{\andname}{\ignorespaces}
\marginsize{2.3cm}{2.3cm}{1.5cm}{1.5cm}
\usepackage{color}
\definecolor{SH}{RGB}{0,0,200}
\definecolor{SH2}{RGB}{0,200,200}

\usepackage{setspace}
\makeatletter

\makeatletter

\begin{document}

\title{Hertz-rate metropolitan quantum teleportation}

\author{Si Shen$^{1}$}
\author{Chen-Zhi Yuan$^{1}$}
\author{Zi-Chang Zhang$^{1}$}
\author{Hao Yu$^{1}$}
\author{Rui-Ming Zhang$^{1}$}
\author{Chuan-Rong Yang$^{1}$}
\author{Hao Li$^{4}$}
\author{Zhen Wang$^{4}$}
\author{You Wang$^{1,3}$}
\author{Guang-Wei Deng$^{1,5}$}
\author{Hai-Zhi Song$^{1,3}$}
\author{Li-Xing You$^{4}$}
\author{Yun-Ru Fan$^{1}$}
\author{Guang-Can Guo$^{1,5}$}
\author{Qiang Zhou$^{1,2,5,\ast}$}

\affiliation{$^1$Institute of Fundamental and Frontier Sciences, University of Electronic Science and Technology of China, Chengdu 610054, P. R. China}
\affiliation{$^2$School of Optoelectronic Science and Engineering, University of Electronic Science and Technology of China, Chengdu 610054, P. R. China}
\affiliation{$^3$Southwest Institute of Technical Physics, Chengdu 610041, P. R. China}
\affiliation{$^4$Shanghai Institute of Microsystem and Information Technology, Chinese Academy of Sciences, Shanghai 200050, P. R. China}
\affiliation{$^5$CAS Key Laboratory of Quantum Information, University of Science and Technology of China, Hefei 230026, P. R. China}
\affiliation{Correspondence and requests for materials should be addressed to QZ (e-mail: $^{\ast}$zhouqiang@uestc.edu.cn).}

\maketitle


\textbf{
	\\Quantum teleportation can transfer an unknown quantum state between distant quantum nodes, which holds great promise in enabling large-scale quantum networks. To advance the full potential of quantum teleportation, quantum states must be faithfully transferred at a high rate over long distance. Despite recent impressive advances, a high-rate quantum teleportation system across metropolitan fiber networks is extremely desired. Here, we demonstrate a quantum teleportation system which transfers quantum states carried by independent photons at a rate of 7.1 ± 0.4 Hz over 64-km-long fiber channel. An average single-photon fidelity of $\geqslant$ 90.6 ± 2.6\% is achieved, which exceeds the maximum fidelity of 2/3 in classical regime. Our result marks an important milestone towards quantum networks and opens the door to exploring quantum entanglement based informatic applications for the future quantum internet.}

\vspace{0.5cm}


\renewcommand\section[1]{
	\textbf{#1}
}
\clearpage
\section{\\Introduction}
\\Quantum teleportation\cite{bennett1993teleporting} enables the ‘disembodied’ transfer of an unknown quantum state to a remote location by using quantum entanglement resource with the help of quantum measurement and classical communication. It lies at the heart of the realization of quantum information technologies such as quantum network\cite{kimble2008quantum,wehner2018quantum,long2022evolutionary} and distributed quantum computation\cite{serafini2006distributed}. Since its initial proposal by Bennett et al. in 1993\cite{bennett1993teleporting}, quantum teleportation has been demonstrated in various platforms, including atomic ensembles\cite{bao2012quantum}, single atoms\cite{nolleke2013efficient}, trapped ions\cite{barrett2004deterministic,riebe2004deterministic}, solid-state quantum systems\cite{reindl2018all}, nuclear magnetic resonance\cite{nielsen1998complete} and quantum optics\cite{bouwmeester1997experimental,boschi1998experimental,marcikic2003long,de2004long,takesue2015quantum,valivarthi2020teleportation,ursin2004quantum,landry2007quantum,jin2010experimental,ma2012quantum,yin2012quantum,sun2016quantum,valivarthi2016quantum,ren2017ground,li2022quantum,xia2017long,braunstein1998teleportation,furusawa1998unconditional,huo2018deterministic,zhao2022real,sychev2018entanglement,bussieres2014quantum}. Teleportation systems based on quantum optics offer a promising avenue towards quantum networks, which can be realized in continuous-variable (CV) and discrete-variable (DV) systems, respectively. For instance, the transfer and retrieval for both coherent states \cite{braunstein1998teleportation,furusawa1998unconditional,huo2018deterministic,zhao2022real} and nonclassical states\cite{sychev2018entanglement}  have been experimentally realized with optical modes in CV systems, providing a method to realize deterministic quantum teleportation. However the distance of CV system is limited to around ten kilometers \cite{huo2018deterministic,zhao2022real}, due to the possible increased fragility with respect to the losses of quantum channels\cite{pirandola2015advances}. For global-scale quantum networks\cite{kimble2008quantum,wehner2018quantum}, the distribution range of quantum states needs to be greatly extended to thousands of kilometers using quantum teleportation in DV systems. Till now, this has been realized with multiple degrees of freedom over several meters to more than one thousand kilometers, from the table-top experiments \cite{bouwmeester1997experimental,boschi1998experimental,marcikic2003long,de2004long,takesue2015quantum,valivarthi2020teleportation} to real-world demonstrations\cite{ursin2004quantum,landry2007quantum,jin2010experimental,ma2012quantum,yin2012quantum,sun2016quantum,valivarthi2016quantum,ren2017ground,li2022quantum}. Especially, by using a low-Earth orbit Micius satellite\cite{lu2022micius}, quantum teleportation over 1200 km has been achieved\cite{ren2017ground,li2022quantum}. Despite impressive results, a high-rate quantum teleportation system has yet to be demonstrated, which is desired for advancing the development of quantum networks.\\
\indent
Here we report an experimental realization of a Hertz-rate quantum teleportation system through fiber over a metropolitan range. Our demonstration relies on a high-performance time-bin entangled quantum light source with a single piece of fiber-pigtailed periodically poled lithium niobate (PPLN) waveguide. The quantum states to be teleported are carried by a weak coherent single-photon source with decoy states. The indistinguishability of photons after prior quantum states distribution through fiber channels is ensured with a fully running feedback system. As an important feature of our demonstration, photonic time-bin qubits are teleported at a rate of 7.1 ± 0.4 Hz over a 64-km-long fiber channel. An average single-photon fidelity of $\geqslant$ 90.6 ± 2.6\% is achieved with the decoy state method. Our implementation establishes an important milestone towards quantum internet.

\section{\\Results}
\begin{figure}[ht]
\centering
\includegraphics[width=16cm]{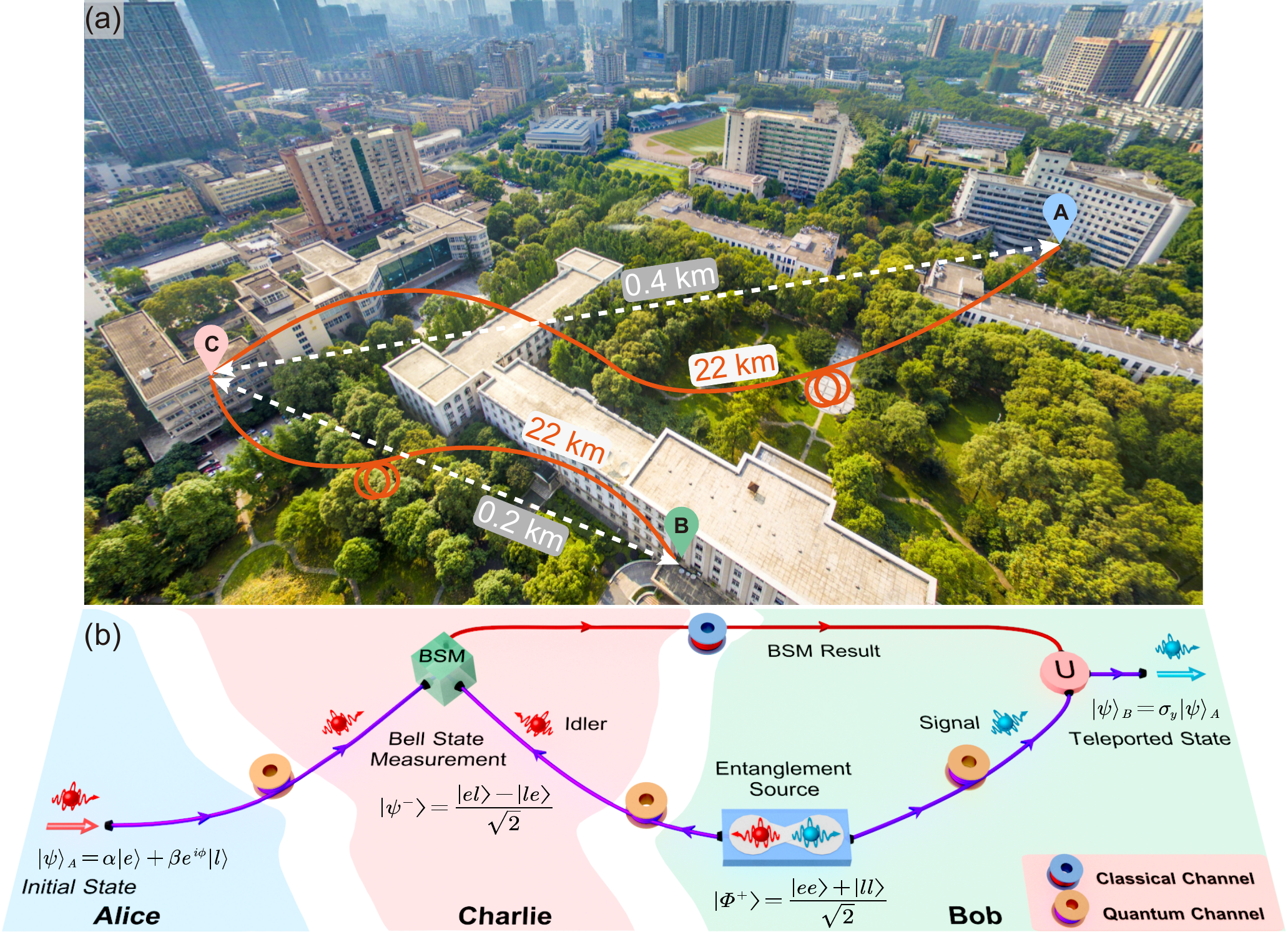}
\caption{\textbf{Three-node quantum teleportation system.} (a) Aerial view of the teleportation system. Alice ‘A’ is located at network’s switching room, Bob ‘B’ and Charlie ‘C’ are located at two separated laboratories. All fibers connecting the three nodes belong to the UESTC backbone network. During the experiment, only the signals created by Alice, Bob and Charlie are transferred through these ‘dark’ fibers. (b) Scheme of the teleportation system. Alice prepares the initial state  $\left| \psi \right> _A$ with a weak coherent single-photon source and sends it to Charlie through a quantum channel ($\mathrm{QC}_{\mathrm{A}\rightarrow \mathrm{C}}$). An entanglement source at Bob generates a pair of entangled photons in the state $\left| \varPhi ^+ \right> $  and then sends the idler photon to Charlie via another quantum channel ($\mathrm{QC}_{\mathrm{B}\rightarrow \mathrm{C}}$). The signal photon is stored in a fiber spool. Charlie implements a joint Bell-state measurement (BSM) between the qubit sent by Alice and Bob, projecting them onto one of the four Bell states $\left| \psi ^- \right>$. Then the BSM result is sent to Bob via a classical channel (CC), who performs a unitary (U) transformation on the signal photon  to recover the initial state (see Methods).}
\label{fig:1}
\end{figure}

\\\textbf{Experimental setup.} Figure \ref{fig:1}(a) shows an aerial photography of the campus of University of Electronic Science and Technology of China (UESTC) indicating the distances between the locations Alice, Bob and Charlie. Figure \ref{fig:1}(b) shows the scheme of our teleportation system, and Fig. \ref{fig:2} depicts its experimental setup. To be compatible with the structure of quantum networks\cite{wei2022towards}, the quantum state to be teleported should be carried by an independent single-photon source, which is different with the Rome scheme\cite{boschi1998experimental}. In our demonstration, the quantum bit (qubit) sender, Alice, located at a switching room of the backbone network of the campus, prepares a weak coherent single-photon source, which is used to encode photonic time-bin qubit, i.e., a single-photon wavepacket in a coherent superposition of two time bins. The time-bin qubit is obtained by passing the single-photon wavepacket through an unbalanced Mach-Zehnder interferometer (UMZI) with path-length difference  $\varDelta \tau $. The time-bin qubit can be written as $\left| \psi \right> _A=\alpha \left| e \right> +\beta e^{i\phi}\left| l \right> $, where $\left| e \right> $ represents the early time bin (i.e., a photon having passed through the short arm of the interferometer);  $\left| l \right> $ is the late time bin (i.e., a photon having passed through the long arm); $\phi$  is a relative phase between   $\left| e \right> $ and $\left| l \right> $, and $\alpha ^2+\beta ^2=1$. Alice sends the created quantum states carried by single-photon wavepackets to Charlie, located at a laboratory at a flight distance of 400 m away, through a quantum channel (QC) of 22 km, i.e., $\mathrm{QC}_{\mathrm{A}\rightarrow \mathrm{C}}$, including 2 km deployed fiber in field and 20 km fiber spool. The quantum information receiver, Bob, located at another laboratory, 210 m from Charlie, shares with Charlie a pair of time-bin entangled photonic qubits in the state of $\left| \varPhi ^+ \right> =2^{-1/2}\left( \left| ee \right> +\left| ll \right> \right)$, with one at 1549.16 nm (idler) and the other at 1531.87 nm (signal). The idler photons are distributed through another 22 km QC to Charlie, i.e., $\mathrm{QC}_{\mathrm{B}\rightarrow \mathrm{C}}$, including 2 km deployed fiber in field and 20 km fiber spool. Charlie performs the joint Bell-state measurement (BSM) between the qubits sent by Alice and Bob, using a 50:50 fiber beam splitter (BS). We select only projections onto the singlet state of $\left| \psi ^- \right> =2^{-1/2}\left( \left| el \right> -\left| le \right> \right)$, which can be realized by the detection of one photon in each output port of BS with a time difference of 625 ps. When a $\left| \psi ^- \right>$  has been successfully detected, the BSM result is sent to Bob over a classical channel (CC) by means of an optical pulse. In this case, the signal photons at Bob (stored in a 20-km-long fiber spool) are projected onto the state of $|\psi \rangle _B=\sigma _y|\psi \rangle _A$, with $\sigma _y$ being a Pauli matrix. The synchronization of the teleportation system is made through the CCs (see Methods). All fiber spools used in our system are non-zero dispersion-shifted single-mode optical fiber (G.655, Yangtze Optical Fibre and Cable). 
\\
\textbf{Prior entanglement distribution.} The property of prior entanglement distribution is measured before performing the BSM of quantum teleportation. In the experiment, we distribute the idler photons through  $\mathrm{QC}_{\mathrm{B}\rightarrow \mathrm{C}}$ to Charlie while the signal photons are held by a 20 km spool of fiber at Bob. The distributed time-bin entanglement property is characterized with the Franson interferometer, with details shown in Supplementary Materials Note S1. The visibilities of two-photon interference fringes are 94.3 ± 0.1\% and 93.5 ± 0.1\%, respectively, as shown in Fig. \ref{fig:3}(a). The error bars of visibilities are calculated by Monte Carlo simulation assuming Poissonian detection statistics. This result indicates that the quantum entanglement property still maintains after being distributed over 42 km fiber channels. Furthermore, it also allows us to ensure that parameters of two UMZIs can be remotely set as the same in our setup, which is a crucial requirement for the quantum teleportation processes.
\\
\textbf{Indistinguishability of photons at Charlie.} Alice’s and Bob’s photons need to be indistinguishable at Charlie for a successful BSM, which is difficult in long distance quantum teleportation. The spatial and spectral indistinguishabilities are ensured by using single-mode fibers and identical fiber Bragg grating (FBG) filters for both photons. The path-length difference and polarization of the photons are stabilized with an active and automatic feedback system (see Methods). The experimental results of indistinguishabilities at Charlie are shown in Figs. \ref{fig:3}(b) and (c). With our fully running feedback system, we measure the Hong-Ou-Mandel (HOM) interference curve \cite{hong1987measurement} with the time-bin qubits from Alice and Bob, respectively. The result given in Fig. \ref{fig:3}(d) shows a HOM-dip with a visibility of 35.3 ± 1.0\% by Gaussian fitting, approaching the upper bound of 40\% between the coherent state and the thermal state, which corresponds to a single-photon indistinguishability of 88.8 ± 2.4\% at Charlie, with details shown in Supplementary Materials Note S2.
\\
\textbf{Quantum teleportation results.} Two classes of quantum states are prepared to be teleported from Alice to Bob: one class contains qubits lying on the equator of the Poincare sphere (coherent superpositions of   $\left| e \right>$ and $\left| l \right>$  with equal amplitudes, $\left| \psi \right> _A=2^{-1/2}\left( \left| e \right> +e^{i\phi}\left| l \right> \right)$, and the other class contains the two poles of the Poincare sphere ($\left| e \right>$ and $\left| l \right>$). For the equatorial states, a successful teleportation implies Bob’s photon to be in a superposition state ($|\psi \rangle _B=\sigma _y|\psi \rangle _A$). Conditional on the successful BSM result from Charlie through CC, we observe sinusoidal curves of three-fold coincidence with visibilities of 61.4 ± 4.0\% and 60.0 ± 3.9\% for two outputs of UMZI2, respectively, as shown in Fig. \ref{fig:3}(e). The maximum value of three-fold coincidence counts is 335 ± 18 for 200 seconds, indicating that a quantum teleportation rate of 7.1 ± 0.4 Hz is achieved excluding an extra measurement losses of 6.25 dB, i.e., 5.30 dB from UMZI2 and 0.95 dB from single-photon detection. With the measured visibilities, the fidelity for the equatorial states can be calculated as $F_{equator}=\left( 1+V \right) /2$, corresponding to a fidelity of 80.4 ± 2.0\% for the equatorial states\cite{de2004long}, which alone can already represent a strong indication of the quantum teleportation. It is worth mentioning that all the visibilities are obtained without subtracting the background noise. With the UMZI1 in Alice removed, we directly prepare $\left| e \right>$ ($\left| l \right>$) state with a single temporal mode and send it to Charlie for BSM. For the measurement, Bob removes the UMZI2 and accumulates three-fold coincidence counts at the corresponding time bins within a coincidence window of 200 ps. The fidelity $F_{e/l}$ can be calculated by $F_{e/l}=R_c/\left( R_c+R_w \right)$ , where $R_c$  and $R_w$ represent the probability of detecting the correct and wrong state in the pole basis, respectively. The measured fidelity for the $|e\rangle$ input state is 92.2±1.0\% and for the  $|l\rangle$ input state 92.4±1.1\%. Assuming that the performance of equatorial states is the same, i.e., $F_+=F_-=F_{+i}=F_{-i}=F_{equator}$, we apply $F_{avg}=\left( 4F_{equator}+F_e+F_l \right) /6$  to obtain an average fidelity of 84.3±1.7\%, which is significantly above the maximum fidelity of 2/3 in classical regime.

Furthermore, we reconstruct the density matrices $\rho$  of the quantum states after teleportation using quantum state tomography (QST) method\cite{james2001measurement}, as described in Note S3 of Supplementary Materials. Four well-defined states ( $\left| e \right>$, $\left| l \right>$, $\left| + \right>$, and $\left| +i \right>$  are created to perform QST in our system. We calculate fidelities of the quantum teleportation by  $F={ }_B\langle\psi|\rho| \psi\rangle_B$ with the expected states ($|\psi \rangle _B$). Figure \ref{fig:4} shows the density matrices of four quantum states after teleportation obtained by QST. The fidelities for all four prepared states are given in Fig. \ref{fig:5}, which exceed the maximum classical value of 2/3. The more decoherence of $|+\rangle$  and $|+i\rangle$ state results from the residual distinguishability of the photons (see Note S2 in Supplementary Materials), which will not cause any effect on $|e\rangle$ and $|l\rangle$ states. This can be improved by further eliminating the distinguishability of photons in all degrees of freedom, i.e., spatial, spectral, temporal, and polarization degrees\cite{rubenok2013real}. The uncertainty of teleportation fidelities is calculated assuming Poissonian detection statistics and using Monte Carlo simulation. The average fidelity $F_{\mathrm{avg}}=\left( 2\left( F_++F_{+i} \right) +F_e+F_l \right) /6$ is 86.4 ± 4.5\%, showing the quantum nature of the disembodied state transfer from Alice to Bob. \\
\indent
It is noted that the classical fidelity bound of 2/3 is only applied when Alice’s initial states carried with genuine single photons, rather than weak coherent states prepared with attenuated laser pulses. Here we utilize the decoy state method (DSM)\cite{lo2005decoy,wang2005beating,ma2005practical} to estimate the performance of our system given that genuine single photons are used\cite{valivarthi2016quantum}. In the experiment, we prepare quantum states $\left| e \right>$, $\left| l \right>$, $\left| + \right>$, and $\left| +i \right>$ with varying the mean photon number per qubit at Alice among three values ( $\mu _{A}^{s}=0.088$, $\mu _{A}^{d}=0.029$  and $\mu _{A}^{v}=0$, where $\mu _{A}^{s}$  , $\mu _{A}^{d}$  and $\mu _{A}^{v}$  are the mean photon numbers of the signal, decoy and vacuum state, respectively) and perform quantum teleportation, with details shown in Tables S. \uppercase\expandafter{\romannumeral3} and \uppercase\expandafter{\romannumeral4} of Supplementary Materials. Based on these results, we calculate the lower bounds of $F_{e/l}^{1}$  and  $F_{+/+i}^{1}$ as shown in Fig. \ref{fig:5}, with $F_{Avg}^{1}$ $\geqslant$ 90.6 ± 2.6\%, which significantly violates the classical bound of 2/3 by more than 9 standard deviations, clearly demonstrating the capability of our system for high-fidelity teleportation. We present an analytical model of our teleportation system\cite{valivarthi2016quantum}, and observe a good quantitative agreement between theory and experiment (see Notes S4 and S5 in Supplementary Materials). Finally, we conclude the key metrics of our teleportation system in Table \ref{table1}, where the state-of-the-art teleportation systems in DV with photonic qubits sent by an independent source are summarized as a comparison. Note that, in Table 1, the state-transfer distance corresponds to the total length of quantum channel between Alice and Bob, while the teleportation distance is defined as the bee-line spatial separation between the location of the BSM station and the signal photon at the time of the BSM projection\cite{valivarthi2016quantum}.
\section{\\Discussion}
\\Metrics for a quantum network are of course the rate, fidelity and distance of quantum teleportation. Although our work has moved one important step closer to high-speed quantum teleportation over a metropolitan area, further increases of teleportation rate in our system could be reached by increasing the repetition rate of system, the efficiencies of SNSPDs and BSM, and using multiple spectral channel\cite{yu2022spectrally}. Further insights into photonic quantum information encoding, the use of multiple degrees of freedom\cite{wang2015quantum} or multiple qubits\cite{zhang2006experimental} will also certainly increase the information capacity of quantum teleportation system based on hyperentanglement Bell-state analysis\cite{sheng2010complete,zhou2015complete}. The deviations of the fidelity from unity in our system are mostly due to multiphoton events of quantum light sources and the remaining distinguishability of the two photons undergoing the BSM. We may replace the SNSPDs with photon-number resolving SNSPDs \cite{madsen2022quantum,stasi2022high} to allow post-selection of multiphoton events. Alternatively, another promising solution to multiphoton events from Alice is applying single quantum emitters that can generate individual photons deterministically\cite{anderson2020quantum}. Further, the indistinguishability between the photons from Alice and Bob could be improved by using narrower FBGs (see Supplementary Materials Note S2). To extend the teleportation distance, the combination of low-Earth-orbit satellite links\cite{ren2017ground,li2022quantum} and quantum repeater architecture\cite{briegel1998quantum,duan2001long} may provide a prospective avenue for the long distances beyond 5000 km or so \cite{simon2017towards}. It is also noted that the signal photons in our system, centered at 1531.87 nm, both in terms of wavelength and spectral width, are compatible with quantum memory in erbium-doped materials\cite{saglamyurek2015quantum,saglamyurek2016multiplexed,wei2022storage}. This, in conjunction with entanglement swapping, constitutes an elementary link of a quantum network, which has been realized recently between two solid states quantum memories\cite{liu2021heralded,lago2021telecom,hermans2022qubit}. \\
\indent
In conclusion, we have demonstrated a quantum teleportation system over metropolitan area, where a 7.1 ± 0.4 Hz teleportation rate is achieved with up to 64 km state-transfer distance. An average fidelity of 86.4 ± 4.5\% is measured using QST. Using the DSM, we obtain an average single-photon fidelity of $\geqslant$ 90.6 ± 2.6\%. Our results are further supported by an analytical model which is consistent with measurements of the quantum teleportation system. Finally, our work establishes the possibility of the high-speed quantum information transmission, which serves as a blueprint for the construction of metropolitan quantum network and eventually towards the global quantum internet.

\section{\\Methods}
\begin{figure}[ht]
\centering
\includegraphics[width=16cm]{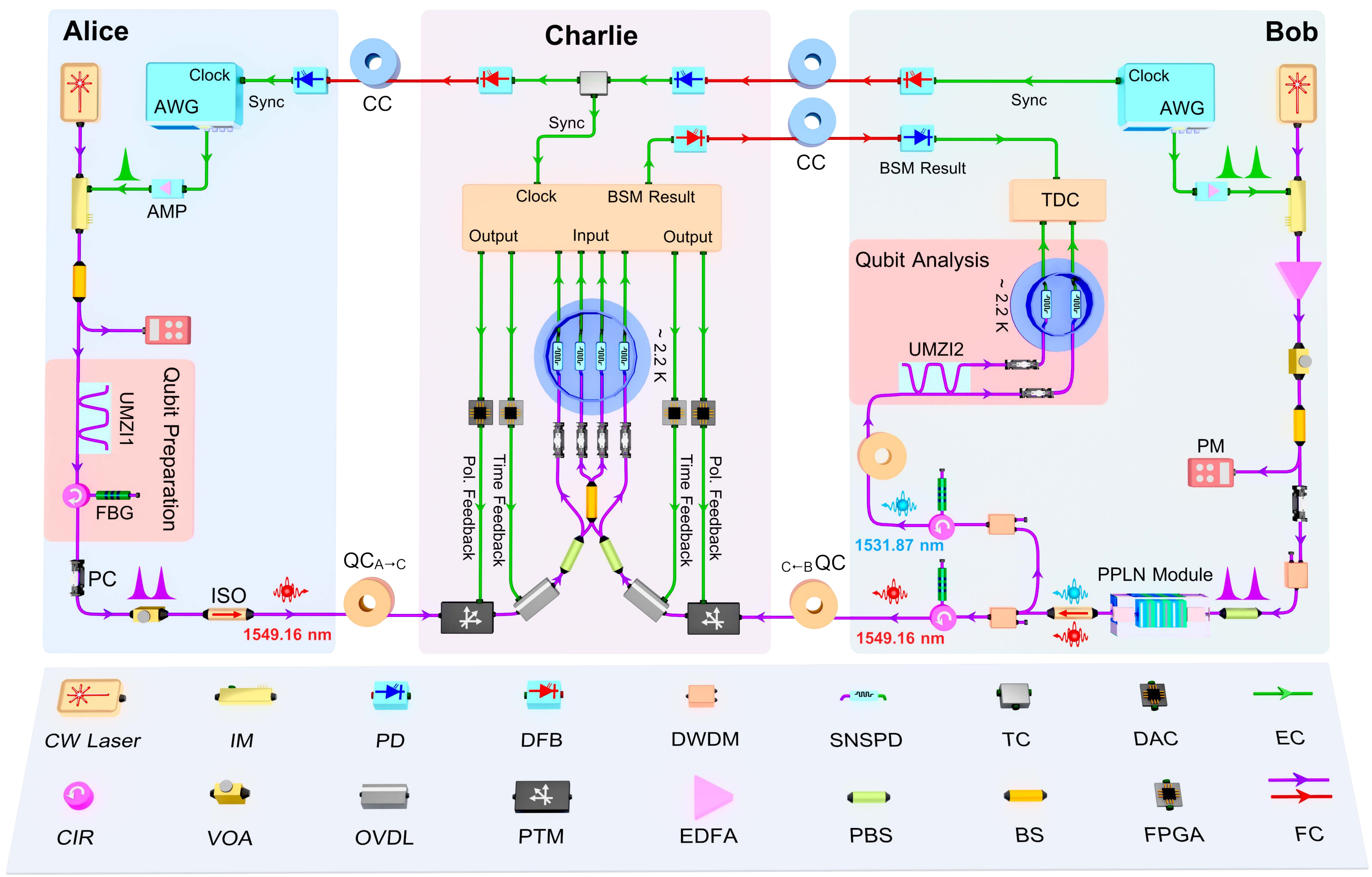}
\caption{\textbf{Experimental setup.} Alice’s setup. The 65-ps-long pulses of light are created by modulating 1549.16 nm continuous wave (CW) laser (PPCL300, PURE Photonics) at 500 MHz rate with an intensity modulator (IM). The drive signal is generated by an arbitrary waveform generator (AWG) and amplified by a 25-GHz-bandwidth amplifier (AMP), synchronized with Bob’s clock through a classical channel (CC, blue line). A fiber beam splitter (BS) with a ratio of 99:1 and a powermeter (PM) are used to monitor the power of the laser pulses. Subsequently, an unbalanced Mach-Zehnder interferometer (UMZI1, MINT, Kylia) with a path-length difference equivalent to 625 ps is applied to prepare the time-bin qubits to be teleported. Following with a spectrally filtering by a 10-GHz-wide fiber Bragg grating (FBG) combined with an optical circulator (CIR) and a strong attenuation to the single photon level by a variable optical attenuator (VOA), the prepared qubits are sent to Charlie through a 22 km fiber quantum channel (QC), $\mathrm{QC}_{\mathrm{A}\rightarrow \mathrm{C}}$ yellow line - featuring 6.8 dB loss. Bob’s setup. Two pump laser pulses separated by 625 ps with the same repetition rate of Alice are generated using a 1540.56 nm CW laser (PPCL300, PURE Photonics) in conjunction with an IM. The pump power is amplified, adjusted, and monitored by an erbium-doped fiber amplifier (EDFA), VOA, and 99:1 BS with a PM, respectively. A polarization controller (PC) and polarization beam splitter (PBS) are used to ensure the polarization alignment for maximizing the efficiency of phase matching in the periodically poled lithium niobate waveguide. The time-bin entangled state of $\left| \varPhi ^+ \right> =2^{-1/2}\left( \left| ee \right> +\left| ll \right> \right)$  is generated using cascaded second-order nonlinear processes in the PPLN waveguide module (see Methods), with mean photon pair number of   $\mu _{\mathrm{SPDC}}=0.042$ in the experiment. The entangled photon pairs are spectrally filtered into signal (1531.87 nm) and idler (1549.16 nm) ones using dense-wavelength division multiplexers (DWDMs) and FBGs with a full width at half maximum bandwidth of 125 GHz and 10 GHz, respectively. The idler photons are sent to Charlie via another 22 km fiber QC, $\mathrm{QC}_{\mathrm{B}\rightarrow \mathrm{C}}$  - featuring 6.4 dB loss and the state of signal photons (stored in a 20 km fiber spool) is analyzed using UMZI2 (625 ps transmission delay, MINT, Kylia), two superconducting nanowire single photon detectors (SNSPDs, P-CS-16, PHOTEC) - cooled to 2.2 K in a cryostat and with 80\% detection efficiency, and a time-}
\label{fig:2}
\end{figure}

\begin{figure}[ht]
\centering
\includegraphics[width=16cm]{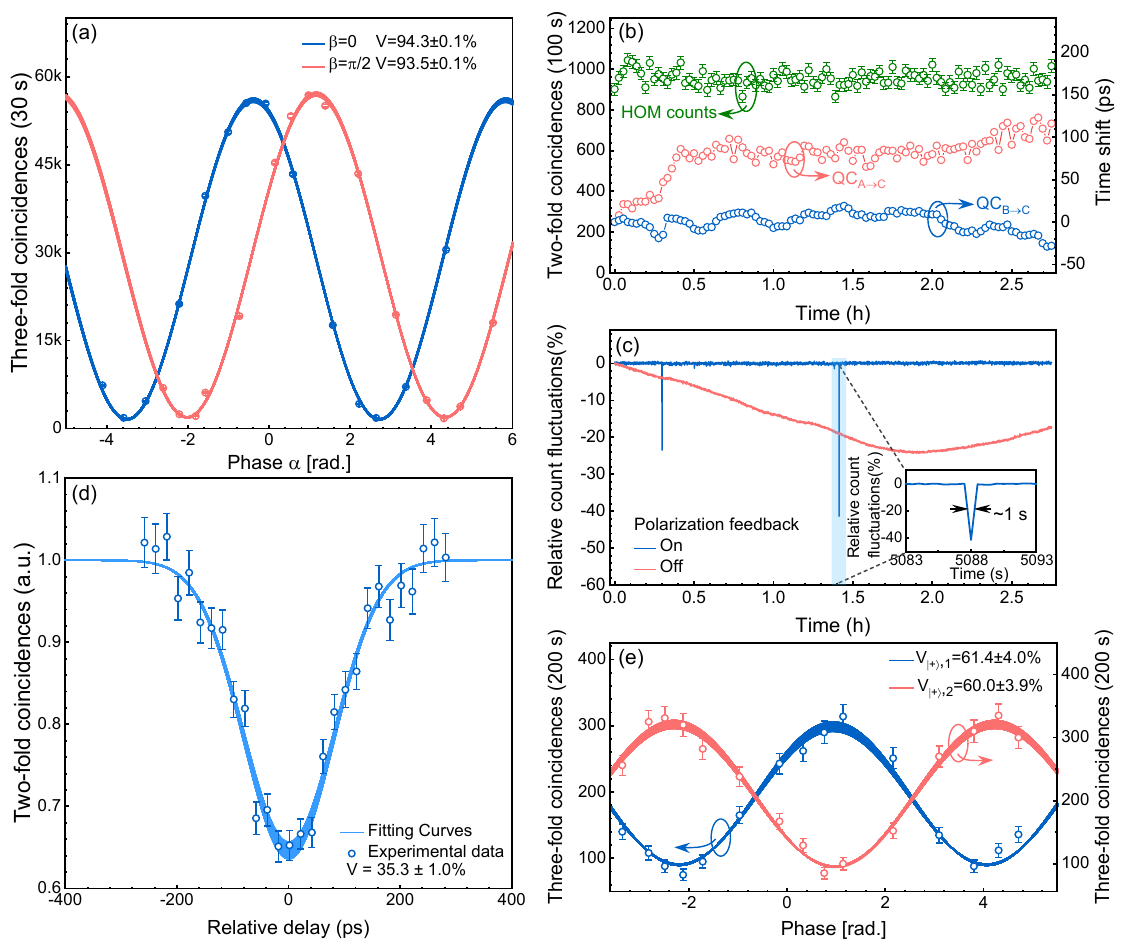}
\caption{\textbf{ Experimental results of prior entanglement distribution, indistinguishability of photons at Charlie, and teleportation of equatorial states.} (a) Two-photon Franson interference fringes of time-bin entanglement source after distribution. Blue and red circles show the coincidence counts for the phase of UMZI1 on idler path set at 0 and $\pi/2$, respectively. The visibilities of the fitting curves are 94.3 ± 0.1\% and 93.5 ± 0.1\%, with the uncertainties calculated using the Monte Carlo method. (b) Automatic timing control on  $\mathrm{QC}_{\mathrm{A}\rightarrow \mathrm{C}}$ and  $\mathrm{QC}_{\mathrm{B}\rightarrow \mathrm{C}}$, respectively. Red (blue) circles represent the drifts of Alice’s (Bob’s) photons arrival times with respect to the system clock. Green circles represent the coincidence counts of HOM interference per 100 seconds with active feedback. (c) Automatic polarization feedback on  $\mathrm{QC}_{\mathrm{A}\rightarrow \mathrm{C}}$. Red (Blue) lines correspond to relative fluctuations of  $\mathrm{QC}_{\mathrm{A}\rightarrow \mathrm{C}}$  with feedback off (on). (d) Normalized HOM interference curve after fiber transmission of photons at Charlie. The visibility of the HOM curve is 35.3 ± 1.0\%, corresponding to a single-photon indistinguishability of 88.8 ± 2.4\% at Charlie, with details shown in Supplementary Materials Note S2. (e) Teleportation of equatorial states. The red and blue circles represent the three-fold coincidence counts from the two outputs of UMZI2 at Bob. The visibilities of the fitting curves are 61.4 ± 4.0\% and 60.0 ± 3.9\%, respectively, indicating the coherence property of Alice’s state is successfully teleported to the signal photons. All error bars are calculated by Monte Carlo simulation assuming Poissonian detection statistics.}
\label{fig:3}
\end{figure}

\begin{figure}[ht]
\centering
\includegraphics[width=12cm]{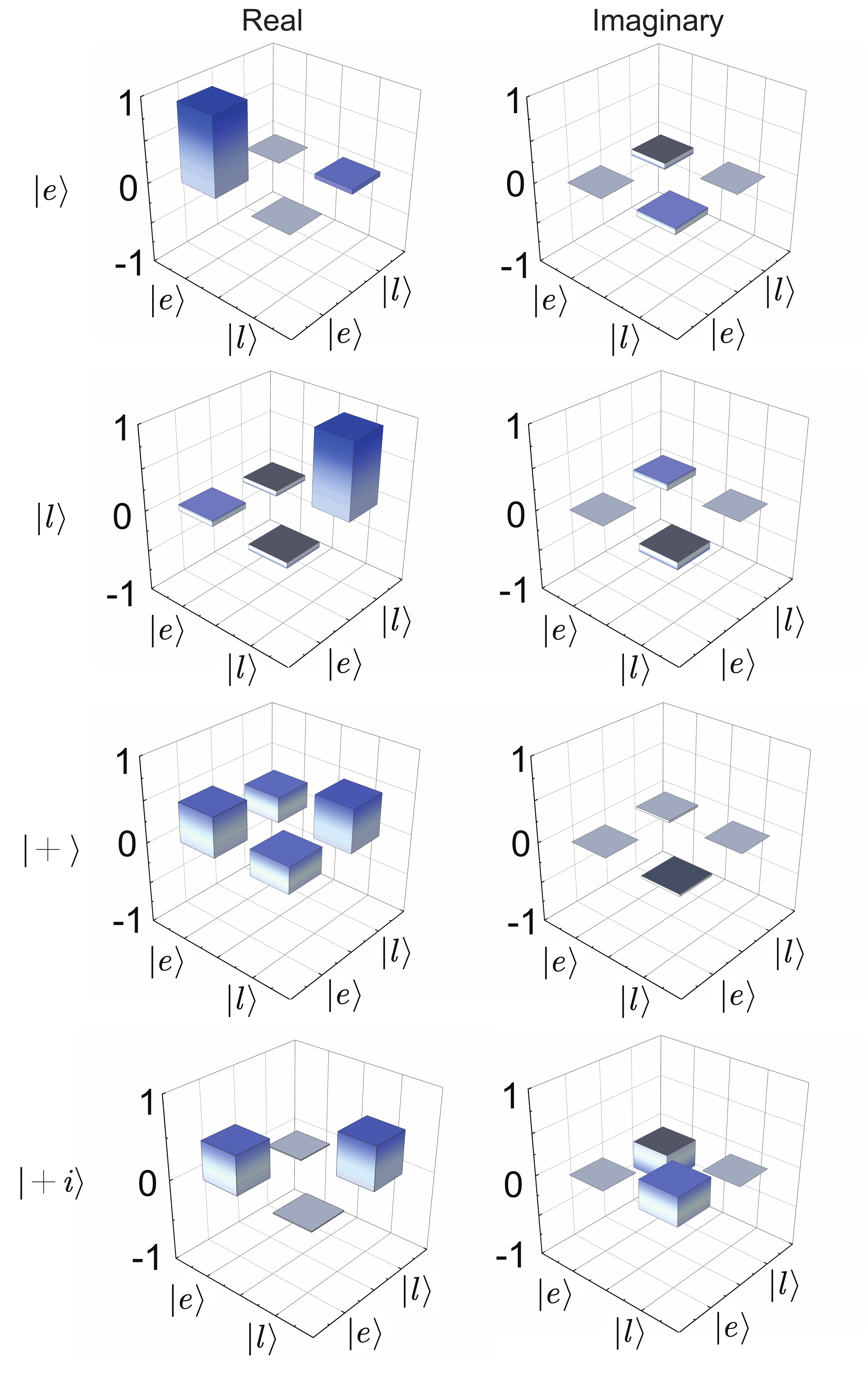}
\caption{\textbf{Density matrices of four quantum states after teleportation.} The real and imaginary parts of the reconstructed density matrices of four different input states prepared at Alice. The state labels denote the states expected after teleportation. The mean photon number per qubit is $\mu _{\mathrm{A}}$ = 0.029 and the mean photon pair number is $\mu _{\mathrm{SPDC}}$ = 0.042.}
\label{fig:4}
\end{figure}

\begin{figure}[ht]
\centering
\includegraphics[width=14cm]{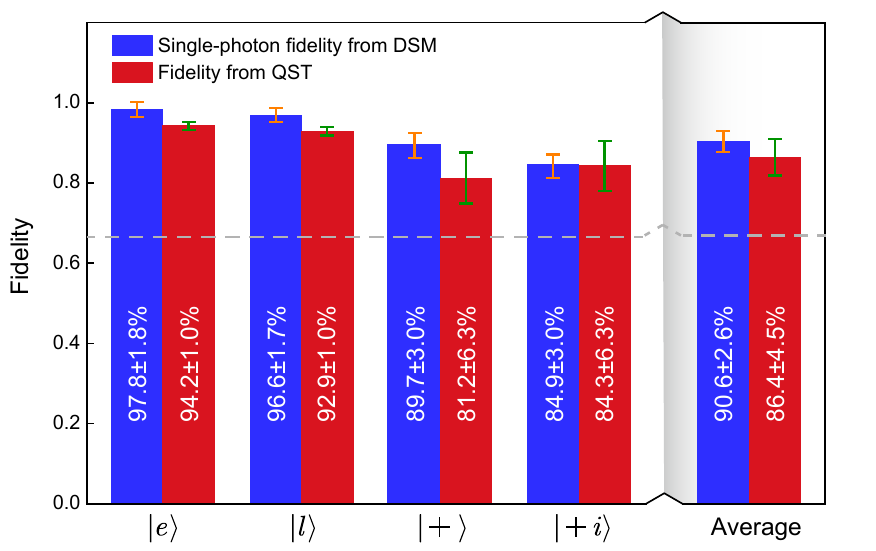}
\caption{\textbf{Individual and average fidelities of four teleported states with ideal state, obtained with quantum state tomography (QST) method and the decoy state method (DSM).} Red bars are fidelities measured using QST. Blue bars are fidelities obtained with DSM. Both fidelities from the two methods exceed the classical limit of 2/3, i.e., the dashed gray line. For the QST and DSM we set $\mu _{\mathrm{SPDC}}$  = 0.042. Error bars are calculated using Monte Carlo simulation, assuming Poissonian detection statistics (see Supplementary Materials Note S5 and Tables S. \uppercase\expandafter{\romannumeral3} and \uppercase\expandafter{\romannumeral4} for more calculation and statistics details).}
\label{fig:5}
\end{figure}

\\
\textbf{PPLN module design.} The entangled photon pairs are generated using cascaded nonlinear processes of second harmonic generation (SHG) and spontaneous parameter down conversion (SPDC) in a periodically poled lithium niobate (PPLN) waveguide module\cite{yu2022spectrally,lefebvre2021compact,zhang2021high}. By fiber-integrating the PPLN waveguide with noise-rejecting filters\cite{zhang2021high}, the spontaneous Raman scattering noise photons generated in the module are greatly reduced, and entangled photon pairs with a high rate under the same coincidence-to-accidental ratio (CAR) are obtained. More details about parameters of our PPLN module are listed in Table S. \uppercase\expandafter{\romannumeral1} of Supplementary Materials.
\\
\textbf{Synchronization.} The master clock of the teleportation system is generated from an AWG at Bob and converted into optical pulses by using distributed feedback (DFB) lasers. The optical pulses are sent through the CCs from Bob to Charlie, and from Charlie to Alice. Charlie and Alice receive the optical pulses and convert them into electrical signals using photon detectors (PDs), the outputs of which are used for synchronization at both stations. As shown in Fig. \ref{fig:2}, Charlie is connected to Alice and Bob through dark fibers of campus backbone networks. Among them, the fiber that carries photonic qubits is referred to as the quantum channel (QC) and the fiber that transmits optical pulses is referred to as the CC. In addition, both unbalanced interferometers at Alice and Bob are calibrated and stabilized by single photon interference. This method permits us to carry out the preparation and measurement of time-bin qubits with stable phases.
\\
\textbf{Stabilization to ensure the indistinguishability of photon.} A successful Bell-state measurement (BSM) relies on the indistinguishability of the two photons, which are generated by independent sources and have been distributed through a 22-km-long fiber channel for each. To do that, the spatial indistinguishability is ensured by using single-mode optical fibers. The spectral indistinguishability of photons from Alice and Bob is ensured by spectral filtering with separate 10-GHz-wide temperature-stabilized FBGs. However, in a real-world quantum teleportation system, the length and birefringence of the optical fiber are influenced by external environments, such as the strain and temperature fluctuations, which make photons distinguishable in degrees of temporal mode and polarization mode. To overcome these challenges, we develop the following experimental techniques to stabilize the QCs (see Supplementary Materials Fig.S. 4 for the schematic).

\emph{Automatic timing control}. We measure the arrival times of Alice’s and Bob’s photon respectively, and compensate their time drifts with respect to the system clock. As shown in Fig. \ref{fig:2}, the detection signals of photons from the reflection port of the polarization beam splitter (PBS) and the system clock are sent to a time-to-digital converter (TDC) to record the arrival times of photons from each channel for every 10 seconds. The drift signals of arrival time in each channel are obtained with a field-programmable gate array (FPGA) circuit. Then the drift signals are fed to two optical variable delay lines (OVDLs, MDL-002, General Photonics) in the two QCs to compensate for the arrival time drifts with a resolution of 1 ps. As shown in Fig. \ref{fig:3}(b), within $\sim$3 hours measurement, a time shift of 120 ps (-40 ps) is applied to compensate timing drift in timing through $\mathrm{QC}_{\mathrm{A}\rightarrow \mathrm{C}}$ ($\mathrm{QC}_{\mathrm{B}\rightarrow \mathrm{C}}$), respectively. During the measurement, the minimum count of the HOM interference is 954 ± 37 per 100 seconds. The result shows that despite the timing drifts in two QCs are much larger than the duration of the single-photon wavepacket, the teleportation should still succeed with our active timing control.

\emph{Automatic polarization control}. At Charlie, the photons from Alice and Bob pass through two PBSs so that the polarization indistinguishability between them is naturally satisfied. However, to ensure the minimum loss of photons through PBS, the polarization must be set and maintained, which can be achieved by an automatic polarization control system to compensate for the polarization drifts. In our experiment, we perform automatic polarization control on both $\mathrm{QC}_{\mathrm{A}\rightarrow \mathrm{C}}$ and $\mathrm{QC}_{\mathrm{B}\rightarrow \mathrm{C}}$, with schematic setup shown in Fig.S. 4. For instance, to control the polarization of $\mathrm{QC}_{\mathrm{A}\rightarrow \mathrm{C}}$, we monitor the detection counts of Alice’s photons from the reflection port of the PBS per 10 seconds. The count number is sent to a digital to analog convertor (DAC) circuit, which generates analog feedback signal. The feedback signal is fed to a polarization track module (PTM, POS-002, General Photonics), which ensures the maximum counts of the transmission port of the PBS by automatic polarization control. As shown in Fig. \ref{fig:3}(c), the average fluctuations of $\mathrm{QC}_{\mathrm{A}\rightarrow \mathrm{C}}$  within $\sim$3 hours are limited to 0.2\% with our automatic polarization feedback (blue line), and to 15.4\% without feedback (red line). The vibrations in the blue line are caused by actively controlling the polarization state of the photons, which recovers within 1 second with our polarization feedback system, as shown in the inset of Fig. \ref{fig:3}(c).
\\
\textbf{Data acquisition.} Charlie performs $\left| \psi ^- \right>$ BSM with photonic qubits to be teleported  from Alice and idler photons (1549.16 nm) from Bob. When two photons arrive on two different detectors with a time delay of 625 ps, a successful $\left| \psi ^- \right> $ detection is obtained. Successful BSM results are transmitted through the CC to Bob by classical optical pulses, which are converted back to electrical signals by using a PD. The signals are then sent to a TDC to perform a three-fold coincidence measurement with detections of stored signal photons (1531.87 nm) from the outputs of UMZI2 at Bob. The time delay between the BSM result and the detection of signal photons is implemented by using a configurable electronic delay module on the TDC.
\section{\\Data availability}\\
The data that support the findings of this study are available from the corresponding author on reasonable request.

\section{\\Acknowledgments}\\
This work was supported by the National Key Research and Development Program of China (Nos. 2018YFA0307400, 2018YFA0306102), National Natural Science Foundation of China (Nos. 61775025, 91836102, U19A2076, 62005039), Innovation Program for Quantum Science and Technology (No. 2021ZD0301702), Sichuan Science and Technology Program (Nos. 2021YFSY0066, 2021YFSY0062, 2021YFSY0063, 2021YFSY0064, 2021YFSY0065). The authors thank X.X.H, Y.X.L and L.B.Z from the Information Center of the University of Electronic Science and Technology of China (UESTC) for providing access to the campus fiber network and for the help during the experiment.

\section{\\Author contributions}\\
 QZ conceived and supervised the project. SS and CY mainly carried out the experiment and collected the experimental data with help of other authors. HL, LY and ZW developed and maintained the SNSPDs used in the experiment. SS, CY and QZ analyzed the data. SS and QZ wrote the manuscript with inputs from all other authors. All authors have given approval for the final version of the manuscript.

\section{\\Competing interests}\\
\\The authors declare that they have no competing interests.

\bibliography{myref}

\clearpage

\clearpage
\clearpage
\begin{spacing}{1.1}
\noindent
to-digital converter (TDC, ID900, ID Quantique). Charlie’s setup. The photons from Alice and Bob are projected onto the $\left| \psi ^- \right> $   Bell state using a 50:50 BS and two SNSPDs with 60\% detection efficiency. To ensure the indistinguishability of the two photons distributed through a 22-km-long fiber channel for each, we actively stabilize the arrival times and polarization with an active and automatic feedback system on both  $\mathrm{QC}_{\mathrm{A}\rightarrow \mathrm{C}}$ and   $\mathrm{QC}_{\mathrm{B}\rightarrow \mathrm{C}}$ channels. The timing and polarization feedback signals (Time Feedback and POL Feedback) are generated from field-programmable gate array (FPGA) circuits and digital to analog convertor (DAC) circuits, respectively, and sent to optical variable delay lines (OVDLs, MDL-002, General Photonics) and polarization tracker modules (PTMs, POS-002, General Photonics) to compensate for the time and polarization drifts. Two optical isolators (ISOs) with ~55 dB isolations are used to shied Alice and Bob from attacks. The synchronization (SYNC) between the three nodes is performed by classical optical pulses through classical channels (CCs), and assisted with AWGs, distributed feedback (DFB) lasers, photon detectors (PDs) and a tee connector (TC). Both QC and CC are dark fiber cables (FC). The electronic cables (EC) are denoted by red lines (see Methods for more details about stabilization and synchronization).
\end{spacing}

\clearpage

\clearpage

\clearpage

\clearpage
\newcommand{\tabincell}[2]{\begin{tabular}{@{}#1@{}}#2\end{tabular}} 
\begin{table}[htbp]\color{black}
	\caption{Comparison between the state-of-the-art results and our work.}
	\label{table1}
	\begin{ruledtabular}
		\begin{tabular}{c@{}c@{}c@{}c@{}c@{}c@{}c@{}}
		Year & \tabincell{c}{State-transfer\\distance (km)} &\tabincell{c}{Teleportation\\distance (km)} & Fidelity & \tabincell{c}{Rate\\(Hz)} & Channel\\
		\hline
		1997\cite{bouwmeester1997experimental} &0.1 & 0.1 & 70\% & $3*10^{-2}$ & Fiber \\
	    2003\cite{marcikic2003long} &2 & 0.1 & 81\% & $5*10^{-2}$ & Fiber \\
	    2004\cite{de2004long} &6 & 0.1 & 78\% & $5*10^{-2}$ & Fiber \\
	    2007\cite{landry2007quantum} &1 & 0.6 & 93\% & $2*10^{-3}$ & Fiber \\
	    2012\cite{ma2012quantum} &143 & 0.1 & 86\% & $3*10^{-2}$ & Free space \\
		2012\cite{yin2012quantum} &97 & 0.1 & 80\% & $8*10^{-2}$ & Free space \\
		2014\cite{bussieres2014quantum} &25 & 0.1 & 81\% & $2*10^{-3}$ & Fiber \\
		2015\cite{takesue2015quantum} &102 & 0.1 & 84\% & $2*10^{-2}$ & Fiber \\
		2016\cite{sun2016quantum} &60 & 6 & 91\% & $5*10^{-4}$ & Fiber \\
		2016\cite{valivarthi2016quantum} &17 & 6 & 80\% & $2*10^{-1}$ & Fiber \\
	    2017\cite{ren2017ground} &1400 & 0.1 & 80\% & $1*10^{-1}$ & Free space \\
		2020\cite{valivarthi2020teleportation} &44 & 0.1 & 89\% & $9*10^{-3}$ & Fiber \\
		\textbf{2022(Our work)} &\textbf{64} & \textbf{0.2} & \textbf{91\%} & \textbf{7.1 ± 0.4} & \textbf{Fiber} \\
		\end{tabular}
	\end{ruledtabular}
\end{table}
\clearpage

\section{\centerline{Supplementary Materials}}
\renewcommand{\baselinestretch}{1.5}
\renewcommand\figurename{Fig.S.}
\renewcommand\tablename{TABLE S.}
\renewcommand{\andname}{\ignorespaces}
\setcounter{figure}{0}
\setcounter{table}{0}
\subsection*{Note S1: Generation and prior distribution of entangled photon pairs}
\begin{figure}[ht]
\centering
\includegraphics[width=16cm]{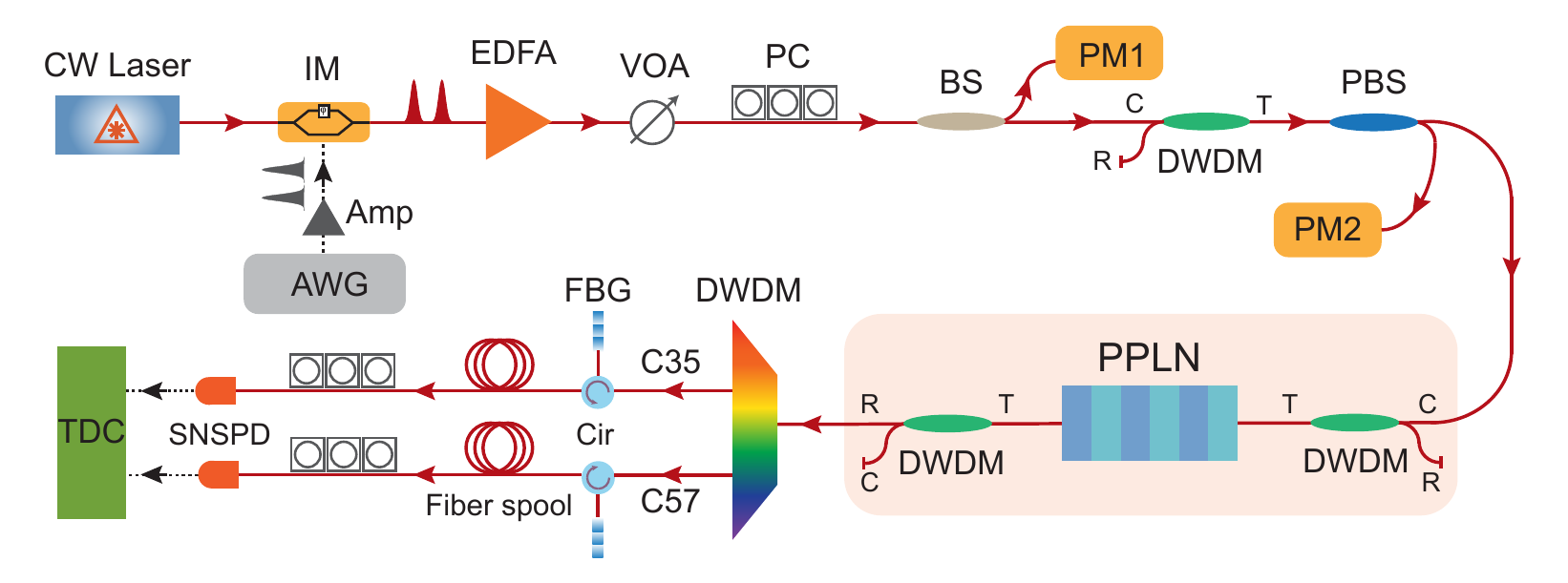}
\caption{\textbf{Experimental setup for generation and characterization of entangled photons pairs.} CW-Laser: continuous wave laser, IM: intensity modulator, EDFA: erbium-doped fiber amplifier, VOA: variable optical attenuator, PC: polarization controller, BS: 99:1 beam splitter, PM: powermeter, DWDM: dense wavelength division multiplexer, PBS: polarization beam splitter, PPLN: periodically poled lithium niobate, Cir: circulator, FBG: fiber Bragg grating, SNSPD: superconducting nanowire single photon detector, TDC: time-to-digital converter.}
\label{figS1}
\end{figure}

\begin{figure}[ht]
\centering
\includegraphics[width=16cm]{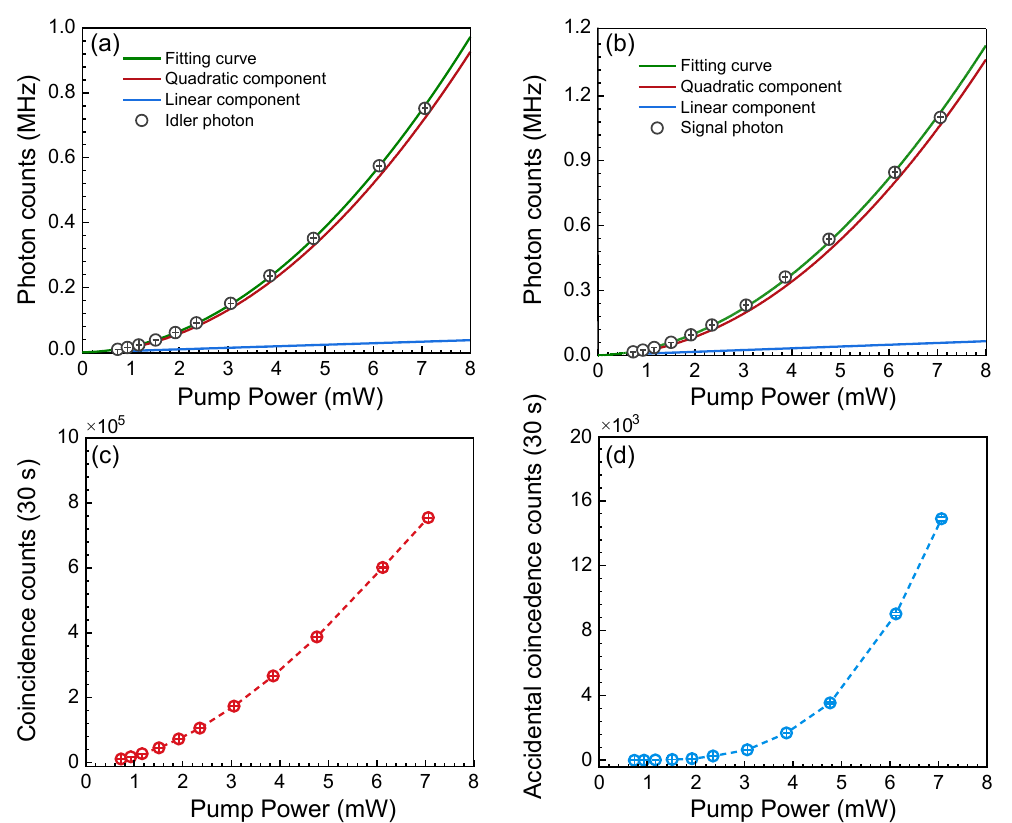}
\caption{\textbf{Properties of entangled photon pairs.} (a) and (b) Idler and signal photon counts versus pump power. The black circles represent measured idler and signal counts with different pump powers. The green line is the quadratic polynomial fitting curve of the measured counts. The quadratic and liner parts are shown as the red and blue lines, respectively. (c) and (d) Coincidence counts and accidental coincidence counts versus pump power. The red circles are measured coincidence counts within a coincidence window of 200 ps for 30 seconds under different pump power levels. The blue circles are measured accidental coincidence counts within a coincidence window of 200 ps for 30 seconds under different pump power levels.}
\label{figS2}
\end{figure}

\noindent
Figure S. \ref{figS1} shows the setup for the generation and characterization of entangled photon pairs. In the experiment, the signal (idler) photon counts $N_s$ ($N_i$) and the coincidence (accidental coincidence) counts  $N_{co}$ ($N_{ac}$) are measured under different pump power levels, which can be expressed as\cite{zhou2009polarization,engin2013photon,zhang2021high}:
\begin{equation}
  \begin{array}{l}
	N_s=\left( R+R_s \right) t_s+n_s\\
	N_i=\left( R+R_i \right) t_i+n_i\\
	N_{co}=Rt_st_i+N_{ac}\\
	N_{ac}=N_sN_i\Delta \tau _{bin}\\
\end{array},
\end{equation}
where $R$ represents the generation rate of photon pairs in periodically poled lithium niobate (PPLN) waveguide, $t_s$ ($t_i$) is the collection efficiency of the signal (idler) wavelength,  $R_s$ ($R_i$ ) is the spontaneous Raman scattering (SpRS) noise photons generated in PPLN module,  $n_s$ ($n_i$ ) is the dark counts of superconducting nanowire single photon detectors (SNSPDs) in signal (idler) channel, $\Delta \tau _{bin}$ is the width of the coincidence window - 200 ps in our experiment. The single-side counts of the signal and idler photons are measured under different pump power levels - black circles in Figs.S. \ref{figS2}(a) and (b). The green lines are quadratic polynomial fitting curves, and the quadratic (red line) and linear (blue line) components present the contribution of entangled photon pairs and noise photons, respectively. We can extract the quadratic terms $Rt_{s,i}$  from the fitting curves of $N_{s,i}$. Then the generation rate of photon pairs $R$ can be calculated via Eq. (1) with the measured $N_{co}$ and $N_{ac}$  under each pump power level. The average number of entangled photon pairs ($\mu _{\mathrm{SPDC}}$) at different pump power levels is obtained by $\mu_{\mathrm{SPDC}}=R/R_{rep}$, where $R_{rep}$ is the repetition rate of entangled photon pairs, i.e., 500 MHz in our experiment. For all measurements in the teleportation experiment, the pump power is set at 6.12 mW, giving a $\mu_{\mathrm{SPDC}}$ = 0.042. Figures S. \ref{figS2}(c) and (d) show the $N_{co}$ and $N_{ac}$  measured with different pump power, respectively.

\begin{figure}[ht]
\centering
\includegraphics[width=16cm]{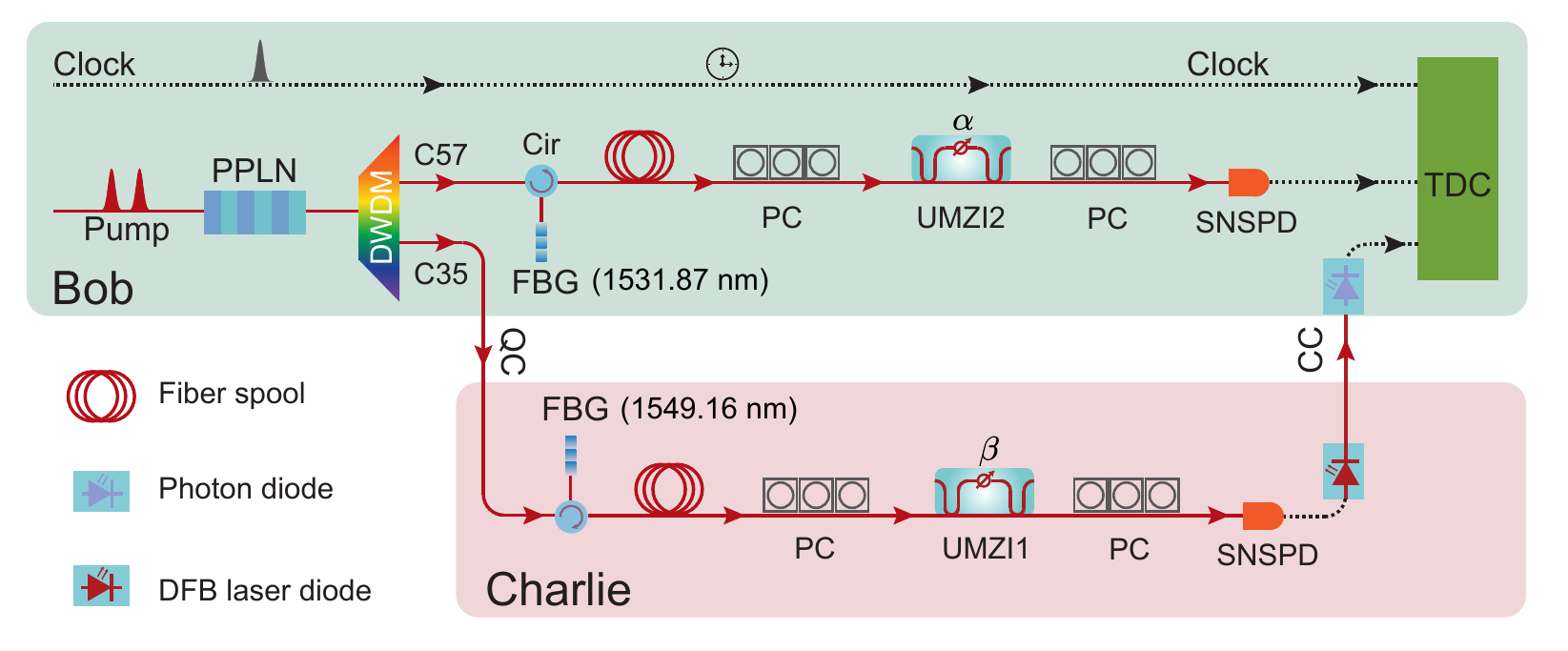}
\caption{\textbf{Schematic for characterization of prior entanglement distribution.} PPLN: periodically poled lithium niobate, DWDM: dense wavelength division multiplexer, Cir: circulator, FBG: fiber Bragg grating, PC: polarization controller, UMZI: unbalanced Mach-Zehnder interferometer, SNSPD: superconducting nanowire single photon detector, TDC: time-to-digital converter, QC: quantum channel, CC: classical channel.}
\label{figS3}
\end{figure}

\indent
Figure S. \ref{figS3} shows the schematic for prior entanglement distribution in our experiment. The length of the fiber spool is about 20 km. In both Charlie and Bob, two unbalanced Mach-Zehnder interferometers (UMZIs, UMZI1 at Charlie and UMZI2 at Bob) and superconducting nanowire single photon detectors (SNSPDs) are used to project the time-bin qubit onto the time or energy bases, where the time delay difference (625 ps) between the long and short arms of UMZIs equals to the interval of time-bins\cite{takesue2009implementation}. Charlie sends the detection signal of 1549.16 nm photons to Bob through the classical channel (CC), who performs three-fold coincidence counts with the system clock and detection signal of 1531.87 nm photons. We obtain two-photon interference fringes when the phase $\beta $  of UMZI1 at Charlie is fixed at 0 and $\pi/2$  and  $\alpha$ of UMZI2 at Bob is scanned, as shown in Fig. 3(a) in the main text.

\subsection*{Note S2: Indistinguishability of photons at Charlie}
\noindent
We apply the Hong-Ou-Mandel (HOM) interference to estimate the single-photon indistinguishability of single-photon wavepackets at Charlie. Let us consider a 50:50 beam splitter (BS) with two input ports and two output ports, with field operators for input and output ports represented as $\hat{a}_1$, $\hat{a}_2$  and $\hat{a}_3$,  $\hat{a}_4$, respectively. We can express these four field operators as:
\begin{equation}
  \begin{aligned}
	\hat{a}_3&=\frac{1}{\sqrt{2}}\left( \hat{a}_1+\hat{a}_2 \right)\\
	\hat{a}_4&=\frac{1}{\sqrt{2}}\left( \hat{a}_1-\hat{a}_2 \right)\\
\end{aligned}.
\end{equation}
The two-photon coincidence probability $P$ with unit indistinguishability can be expressed\cite{ou1988quantum,li2008observation}:
\begin{equation}
  \begin{aligned}
	P\propto &\left. \langle \hat{a}_{3}^{\dagger}(t)\hat{a}_{4}^{\dagger}(t^{'}) \hat{a}_4(t^{'}) \hat{a}_3(t) \right. \rangle\\
	=&\frac{1}{4}\left( \langle \hat{a}_{1}^{\dagger}(t)\hat{a}_{1}^{\dagger}(t^{'}) \hat{a}_1(t^{'}) \hat{a}_1(t) \right. \rangle +\left. \langle \hat{a}_{2}^{\dagger}(t)\hat{a}_{2}^{\dagger}(t^{'}) \hat{a}_2(t^{'}) \hat{a}_2(t) \right. \rangle\\
	&+\left. \langle \hat{a}_{1}^{\dagger}(t)\hat{a}_{2}^{\dagger}(t^{'}) \hat{a}_2(t^{'}) \hat{a}_1(t) \right. \rangle +\left. \langle \hat{a}_{2}^{\dagger}(t)\hat{a}_{1}^{\dagger}(t^{'}) \hat{a}_1(t^{'}) \hat{a}_2(t) \right. \rangle\\
	&\left. -\left. \langle \hat{a}_{2}^{\dagger}(t)\hat{a}_{1}^{\dagger}(t^{'}) \hat{a}_2(t^{'}) \hat{a}_1(t) \right. \rangle -\left. \langle \hat{a}_{1}^{\dagger}(t)\hat{a}_{2}^{\dagger}(t^{'}) \hat{a}_1(t^{'}) \hat{a}_2(t) \right. \rangle \right)\\
	=&\frac{1}{4}\left( g_{1}^{(2)}(0)\bar{n}_{1}^{2}+g_{2}^{(2)}(0)\bar{n}_{2}^{2}+2\bar{n}_1\bar{n}_2-2\bar{n}_1\bar{n}_2I_{12} \right)\\
\end{aligned},
\end{equation}
where  $\bar{n}_1$ and $\bar{n}_2$  are the mean photon number for the two input fields of $\hat{a}_1$ and $\hat{a}_2$, respectively. $g_{1}^{(2)}(0)$  and  $g_{2}^{(2)}(0)$ are second order auto-correlation values for two input fields of $\hat{a}_1$ and $\hat{a}_2$, respectively. $I_{12}\left. =\langle \hat{a}_{1}^{\dagger}(t)\hat{a}_{2}^{\dagger}(t^{'}) \hat{a}_1(t^{'}) \hat{a}_2(t) \right. \rangle /\bar{n}_1\bar{n}_2$  represents the interference term, whose value depends on the overlap between the two input fields. When two input fields are completely overlapped with  $t=t^{'}$, the interference term $I_{12}$  = 0 corresponds to the dip of HOM curve. When two input fields are temporally separated, the interference term $I_{12}$  = 1 corresponds to the wing of HOM curve.
\\
\indent
Assuming the input fields $\hat{a}_1$ and $\hat{a}_2$ are ideal thermal and coherent field, respectively,  $g_{1}^{(2)}(0)$ = 2 and $g_{2}^{(2)}(0)$  = 1 are obtained. We can derive the visibility of HOM interference from Eq. (3):
\begin{equation}
V=\frac{2}{2 \frac{\bar{n}_1}{\bar{n}_2}+\frac{\bar{n}_2}{\bar{n}_1}+2}.
\end{equation}

\indent
Supposing the mean photon number for the two input fields are identical with $\bar{n}_1=\bar{n}_2$, a theoretical upper bound of HOM interference visibility  $V_{theory}$ = 40\% is obtained. The HOM visibility measured in our experiment is $V_{exp}$ = 35.3 ± 1.0\% with identical mean photon number from Alice and Bob (see Fig. 3(d) in the main text), which indicates that the single-photon indistinguishability of single-photon wavepacket at Charlie is $V_{exp}/V_{theory}$  = 88.8 ± 2.4\%, i.e., a residual distinguishability of 11.2±2.4\%. The measured second order auto-correlation value of the idler photons (1549.16 nm) after 10 GHz spectral filtering is 1.88 ± 0.04. Thus we argue that the reduction in the indistinguishability from the ideal value of 1 could be attributed to the imperfection of spectral shape of the filter, which can be further improved.\\
\indent
By using Eq. (4), we calculate the predicted HOM visibility as a function of the mean photon number of teleported qubits $\mu _{\mathrm{A}}$, shown by the blue line in Fig.S. \ref{figS5}(a). The red circles in Fig.S. \ref{figS5}(a) shows the experimental HOM visibility versus  $\mu _{\mathrm{A}}$, which agrees well with the calculation results. For all the calculations and measurements, the mean photon pair number  $\mu _{\mathrm{SPDC}}$ is 0.042.

\subsection*{Note S3: Tomography of teleported state}
\noindent
We use quantum state tomography (QST) to reconstruct the density matrix of the quantum state after teleportation, and then calculate the quantum teleportation fidelity. Based on the scheme in Ref.\cite{james2001measurement}, the density matrix of a single time-bin qubit can be represented by Stokes parameters:
\begin{equation}
  \begin{aligned}
\hat{\rho}=\frac{1}{2}\sum_{i=0}^3{S_i}\sigma _i,
\end{aligned}
\end{equation}
where $\sigma_i$  is the Pauli matrix and  $S_i$ represents the Stokes parameter. Stokes parameters can be calculated by projecting the states to the basis of $\left. |e \right> \left< e| \right.$,  $\left. |l \right> \left< l| \right.$,  $\left. |+ \right> \left< +| \right.$,  $\left. |- \right> \left< -| \right.$,   $\left. |+i \right> \left< +i| \right.$ and  $\left. |-i \right> \left< -i| \right.$, where  $\left. |+ \right>$,  $\left. |- \right>$,  $\left. |+i \right>$ and   $\left. |-i \right>$ are represented by a linear combination of  $\left. |e \right>$ and $\left. |l \right>$:
\begin{equation}
  \begin{array}{l}
	\left. |+ \right> =\left( \left. \left. |e \right> +|l \right> \right) /\sqrt{2}\\
	\left. |- \right> =\left( \left. \left. |e \right> -|l \right> \right) /\sqrt{2}\\
	\left. |+i \right> =\left( \left. \left. |e \right> +i|l \right> \right) /\sqrt{2}\\
	\left. |-i \right> =\left( \left. \left. |e \right> -i|l \right> \right) /\sqrt{2}\\
\end{array}.
\end{equation}
\indent
The projection measurement counts are  $N_e$,  $N_l$,  $N_+$,  $N_-$,   $N_{+i}$ and  $N_{-i}$, respectively. Using these counts, we obtain the Stokes parameters as follows:
\begin{equation}
  \begin{aligned}
	S_0&=N_e+N_l\\
	S_1&=N_+-N_-\\
	S_2&=N_{+i}-N_{-i}\\
	S_3&=N_e-N_l\\
\end{aligned}.
\end{equation}
\indent
Substituting Eq. (7) into Eq. (5), the density matrix $\rho$  after teleportation can be obtained. The teleportation fidelity is calculated with the expected state $\left. |\psi \right>$  by:
\begin{equation}
  F=\left< \psi |\rho |\psi \right>.
\end{equation}


\subsection*{Note S4: Analytical model of teleportation system}
\noindent
We apply the analytical model in Ref.\cite{valivarthi2016quantum} to figure out the main parameters in our experiment, thus improving the performance of our teleportation system. In this method, the sum of the fidelity and the error rate is equal to 1. Hence, the fidelity of the teleported state can be predicted with the probability of three-fold coincidence counts for successful teleportation ($P_S$) and that for failure teleportation ($P_F$)
\begin{equation}
  \begin{aligned}
F=\frac{P_S}{P_S+P_F}
\end{aligned}.
\end{equation}
Notice that in the following experiments, all the measured fidelities are calculated by using Eq. (9), with  $P_S$ obtained by the maximum three-fold coincidence counts and $P_F$  obtained by the minimum three-fold coincidence counts.

Since the average photon number per qubit in our system is much less than 1, we ignore the contribution of higher-order terms to the predicted results. Here, we only consider the following cases: $n_{\mathrm{A}}\leqslant 2, n_{\mathrm{i}}\leqslant 2, n_{\mathrm{A}}+n_{\mathrm{i}}\leqslant 2$  and $n_{\mathrm{s}}\leqslant 2$ ( $n_{\mathrm{A}}$ and   $n_{\mathrm{i}}$ denote the number of photons arriving at the BS from Alice and Bob. $n_{\mathrm{s}}$  is the signal photon number generated by entangled photon pairs). The three-fold coincidence counts probability per qubit $P\left( n_{\mathrm{A}},n_{\mathrm{i}},n_{\mathrm{s}} \right) $  for different cases can be expressed by:
\begin{equation}
\begin{aligned}
	P\left( 1,1,1 \right) &=\frac{1}{4}\mu _{\mathrm{SPDC}}\mu _{\mathrm{A}}\eta _{\mathrm{A}}e^{-\mu _{\mathrm{A}}\eta _{\mathrm{A}}}\eta _{\mathrm{i}}\eta _{\mathrm{s}}\xi _{\mathrm{BSM}}^{2}\xi _{\mathrm{s}}\\
	P\left( 0,2,2 \right) &=\frac{1}{4}\mu _{\mathrm{SPDC}}^{2}e^{-\mu _{\mathrm{A}}\eta _{\mathrm{A}}}\eta _{\mathrm{i}}^{2}\xi _{\mathrm{BSM}}^{2}\left( 1-\left( 1-\xi _{\mathrm{s}}\xi _{\mathrm{s}} \right) ^2 \right)\\
	P\left( 2,0,1 \right) &=\frac{1}{4}\mu _{\mathrm{SPDC}}\left( \mu _{\mathrm{A}}\eta _{\mathrm{A}} \right) ^2\frac{e^{-\mu _{\mathrm{A}}\eta _{\mathrm{A}}}}{2}\left( 1-\eta _{\mathrm{i}} \right) \xi _{\mathrm{BSM}}^{2}\eta _{\mathrm{s}}\xi _{\mathrm{s}}\\
	P\left( 1,1,2 \right) &=\frac{1}{2}\mu _{\mathrm{SPDC}}^{2}\mu _{\mathrm{A}}\eta _{\mathrm{A}}e^{-\mu _{\mathrm{A}}\eta _{\mathrm{A}}}\left( 1-\eta _{\mathrm{i}} \right) \eta _{\mathrm{i}}\xi _{\mathrm{BSM}}^{2}\left( 1-\left( 1-\eta _{\mathrm{s}}\xi _{\mathrm{s}} \right) ^2 \right)\\
\end{aligned},
\end{equation}
where $\mu _{\mathrm{SPDC}}$  represents the average entangled photon pair number per qubit, $\mu _{\mathrm{A}}$  is the average photon number per qubit at Alice,  $\eta _{\mathrm{A}}$ is the transmission probability of quantum channel from Alice to Charlie ($\mathrm{QC}_{\mathrm{A}\rightarrow \mathrm{C}}$),  $\eta _{\mathrm{i}}$ represent the transmission probability of quantum channel from Bob to Charlie ($\mathrm{QC}_{\mathrm{B}\rightarrow \mathrm{C}}$),    $\eta _{\mathrm{s}}$ is the transmission probability of signal photons (stored in a fiber spool),  $\xi _{\mathrm{BSM}}$ is the detection efficiency of SNSPDs used for BSM,  $\xi _{\mathrm{s}}$ is the detection efficiency of SNSPDs for signal photons. All of the experimental parameters in the teleportation system are listed in Table S. \ref{tables2}.\\
\indent
The teleportation fidelity of input states on the equator of the Bloch sphere is given by:
\begin{equation}
F_{+/-}=\frac{1}{2}+\frac{\zeta \left[ P\left( 1,1,1 \right) +P\left( 1,1,2 \right) \right]}{2\left[ P\left( 1,1,1 \right) +P\left( 1,1,2 \right) +P\left( 0,2,2 \right) +P\left( 2,0,1 \right) \right]},
\end{equation}
where $\zeta$  represents the degree of indistinguishability between photons from Alice and Bob (see Note S2). The teleportation rate can be expressed as:
\begin{equation}
N=R_{rep}\times \left[ P\left( 1,1,1 \right) +P\left( 1,1,2 \right) +P\left( 0,2,2 \right) +P\left( 2,0,1 \right) \right],
\end{equation}
where $R_{rep}$  represents the repetition rate of teleportation system. 
From Eqs. (10) and (11), we predict the fidelity of $\left| + \right> $  state as a function of $\mu _{\mathrm{A}}$  - the blue curve in Fig.S. \ref{figS5}(b). An increase of the fidelity with $\mu _{\mathrm{A}}$  is observed. The fidelity reaches a maximum value when the probabilities of receiving one photon from Alice and Bob are equal at Charlie\cite{rarity2005non}. With further increasing of $\mu _{\mathrm{A}}$, the multiphoton events decrease the fidelity. The teleportation rate predicted by our model as a function of state-transfer distance is shown in Fig.S. \ref{figS5}(c). The teleportation rate decays exponentially as the state-transfer distance increases with the length of fiber spool at Bob. Figure S. \ref{figS5}(d) shows the predicted fidelity of  $\left| + \right> $ state with different state-transfer distances, which remains unchanged as the distance increases. As shown in Fig.S. \ref{figS5}(e), we plot the fidelity of  $\left| + \right> $ state as a function of $\mu _{\mathrm{SPDC}}$  with $\mu _{\mathrm{A}}$  = 0.029. The fidelity decreases with $\mu _{\mathrm{SPDC}}$  due to the multiphoton events from the entangled photon pairs. To verify the prediction of our model, we carry out experiments under different situations. Red circles in Fig.S. \ref{figS5}(b) show the measured fidelities of $\left| + \right> $  state with different $\mu _{\mathrm{A}}$. The fidelities of $\left| + \right> $  state and teleportation rates with different state-transfer distances of 44, 64 and 84 km are illustrated by red circles in Figs.S. \ref{figS5}(c) and (d), respectively. The measured fidelities of  $\left| + \right> $ states with different $\mu _{\mathrm{SPDC}}$  are shown by the red circles in Fig.S. \ref{figS5}(e). From the above results, we observe an excellent quantitative agreement between experiments and model.

\subsection*{Note S5: Predicting the fidelity of genuine single photon with decoy state method}
\noindent
The decoy state method (DSM) is originally put forward to defend against photon number splitting attack in quantum key distribution by preparing sender’s source with multiple intensity levels\cite{lo2005decoy,wang2005beating,ma2005practical}.We apply this method and follow Ref.\cite{valivarthi2016quantum} to extract the fidelity and rate of single photon in our teleportation system. In the experiment, the teleported states at Alice are prepared by attenuated laser pulses with different average photon numbers (denoted as signal state $\mu _{A}^{s}$, decoy state  $\mu _{A}^{d}$, and vacuum state  $\mu _{A}^{v}$). The error rate $E^{(1)}$  for the single photon component of the weak coherent single-photon source is upper bounded by\cite{ma2005practical}
\begin{equation}
E^{(1)}\leqslant E_{\mathrm{U}}^{(1)}=\frac{E^{\left( \mu _{A}^{d} \right)}Q^{\left( \mu _{A}^{d} \right)}e^{\mu _{A}^{d}}-E^{(0)}Y^{(0)}}{\mu _{A}^{d}Y_{\mathrm{L}}^{(1)}}.
\end{equation}	 	
$E^{\left( \mu _{A}^{d} \right)}$ and $E^{(0)}$  are the error rates when Alice’s teleported state is encoded in the decoy state and vacuum state, respectively.  $Q^{\left( \mu _{A}^{d} \right)}$ is the corresponding gain, i.e., the probability for three-fold coincidence counts when a weak coherent state with mean photon number  $\mu _{A}^{d}$ is prepared at Alice.  $Y^{(0)}$ is the yield for a vacuum state, and  $Y_{\mathrm{L}}^{(1)}$ is the lower bound of yield for the genuine single photon state. From Eq. (13), we can estimate the upper bound of the error rate of the teleported state prepared using genuine single photons. The parameters except $Y_{\mathrm{L}}^{(1)}$  in Eq. (13) can be measured experimentally. Assuming a weak coherent state is created at Alice with a mean photon number of $\mu$, the corresponding gain can be expressed as:
\begin{equation}
Q^{(\mu )}=\sum_{n=0}^{\infty}{\frac{Y^{(n)}\mu ^ne^{-\mu}}{n!}},
\end{equation}	 	
where $n$ is the photon number of the quantum state,   $Y^{(n)}$ is the yield of an n-photon state, which cannot be measured directly from the experiment, with an exception of $Y^{(0)}$. From Ref.\cite{sun2016quantum},  $Y_{\mathrm{L}}^{(1)}$ can be calculated by:
\begin{equation}
Y^{(1)}\geqslant Y_{\mathrm{L}}^{(1)}=\frac{\mu _{A}^{s}}{\mu _{A}^{s}\mu _{A}^{d}-\left( \mu _{A}^{d} \right) ^2}\left( \begin{array}{c}
	Q^{\left( \mu _{A}^{d} \right)}e^{\mu _{A}^{d}}-\\
\end{array} \right. 
\\
\frac{\left( \mu _{A}^{d} \right) ^2}{\left( \mu _{A}^{s} \right) ^2}Q^{\left( \mu _{A}^{s} \right)}e^{\mu _{A}^{s}}-\frac{\left( \mu _{A}^{s} \right) ^2-\left( \mu _{A}^{d} \right) ^2}{\left( \mu _{A}^{s} \right) ^2}\left. Y^{(0)} \right),
\end{equation}
where  $Q^{\left( \mu _{A}^{s} \right)}$ is the gain of the signal state. With measured  $Q^{\left( \mu _{A}^{s} \right)}$,  $Q^{\left( \mu _{A}^{d} \right)}$,  $Y^{(0)}$,  $E^{\left( \mu _{A}^{d} \right)}$,  $E^{(0)}$,   $\mu _{A}^{s}$ and  $\mu _{A}^{d}$, the upper bound of the error rate  $E_{\mathrm{U}}^{(1)}$ can be obtained from Eqs. (13) and (15). Therefore, we can calculate the lower bound of the teleportation fidelity for quantum states using a genuine single photon source:
\begin{equation}
F^{(1)}=1-E^{(1)}\geqslant 1-E_{\mathrm{U}}^{(1)}\equiv F_{\mathrm{L}}^{(1)}.
\end{equation}
\indent
Based on these results, we obtain an average single-photon fidelity of $\geqslant$ 90.6 ± 2.6\%, with gains and fidelities shown in Tables S. \ref{tables3} and \ref{tables4}, respectively. Considering only the $n$ = 1 term in Eq. (14), the corresponding gain contributed by a single photon input is obtained by $e^{-\mu _{A}^{d}}\mu _{A}^{d}Y_{\mathrm{L}}^{(1)}$ , resulting in a single-photon teleportation rate of $\geqslant$ 6.1 ± 0.7 Hz in our system.
Furthermore, we change the average entangled photon pair number ($\mu _{\mathrm{SPDC}}
$) and measure the fidelities of $|+\rangle $ state with DSM. The green circles in Fig.S. \ref{figS5}(e) show the fidelities of $|+\rangle $ state remain unchanged with $\mu _{\mathrm{SPDC}}
$  increasing (see Tables S. \ref{tables5} and \ref{tables6} for more details).

\clearpage

\clearpage

\clearpage

\clearpage
\begin{figure}[ht]
\centering
\includegraphics[width=16cm]{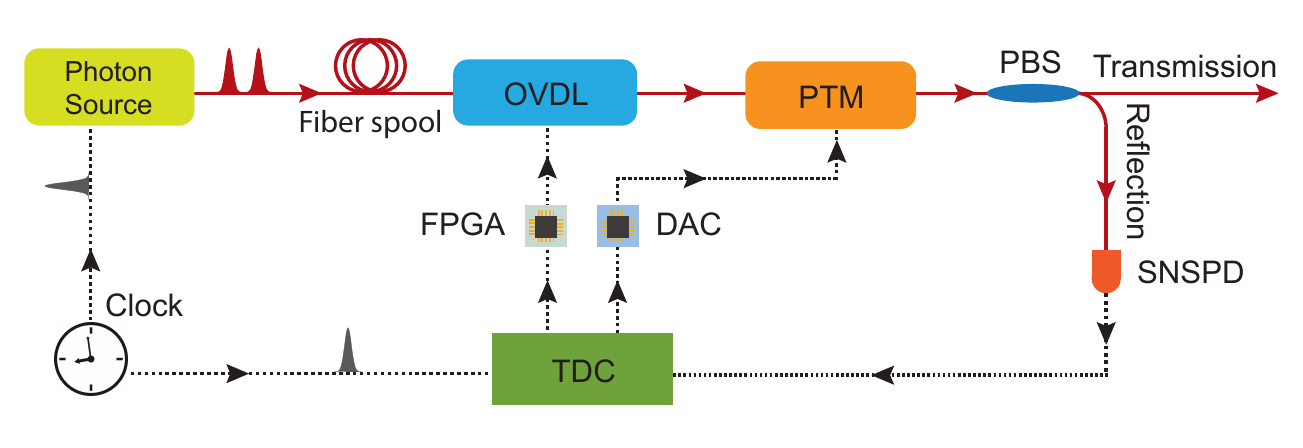}
\caption{\textbf{Schematic for automatic timing and polarization control.} Photon Source: photons from Alice/Bob, OVDL: optical variable delay line, PTM: polarization track module, FPFA: field-programmable gate array, DAC: digital to analog convertor, PBS: polarization beam splitter, SNSPD: superconducting nanowire single photon detector, TDC: time-to-digital converter.}
\label{figS4}
\end{figure}

\clearpage
\begin{figure}[ht]
\centering
\includegraphics[width=16cm]{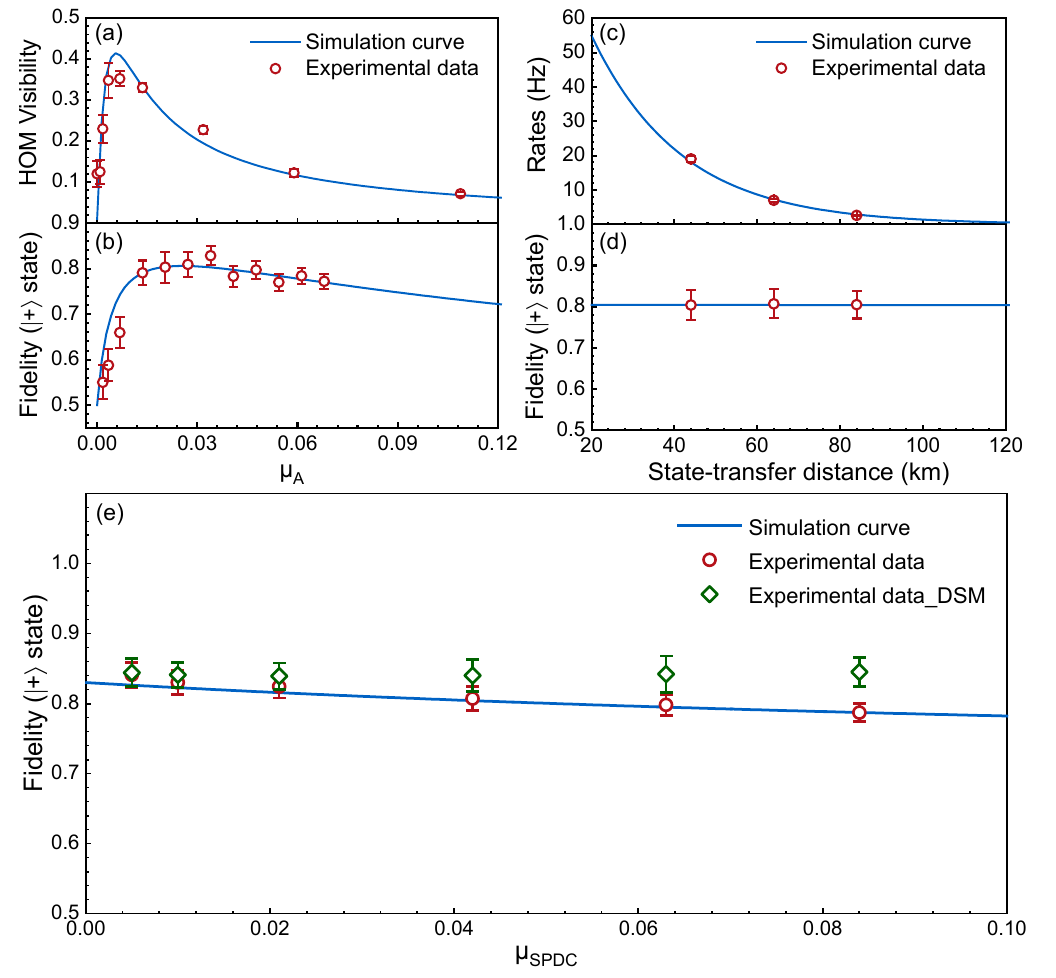}
\caption{\textbf{Calculated result and experimental data of HOM visibilities, teleportation fidelities of $\left| + \right>$ state and teleportation rates.} (a) and (b) Calculated and measured HOM visibilities and teleportation fidelities of $\left| + \right>$  state with different mean photon numbers ($\mu _A$) per qubit at Alice. For all the measurements, $\mu _{\mathrm{SPDC}}$  = 0.042. (c) and (d) Calculated and measured teleportation rates (maximum three-fold coincidence counts) and teleportation fidelities of $\left| + \right>$  state with different state-transfer distances. (e) Calculated and measured teleportation fidelities of  $\left| + \right>$ state with different mean photon pair numbers. Notice that, the circles and lines in (a-e) correspond to the experimental data and calculated results, respectively.}
\label{figS5}
\end{figure}

\clearpage
\begin{table}\color{black}
\centering
	\caption{Parameters of PPLN module.}
	\label{tables1}
\begin{tabular}{l l c}
\hline
\hline
 Type of waveguide & & RPE waveguide \\
 Length of waveguide & & 50 mm \\
 QPM period & & 19 $\mu$m \\
 SHG normalized conversion efficiency & & 500\%/W@1540.56 nm\\
 Length of pigtail & & 20 cm \\
 Input coupling efficiency of PPLN waveguide & & 73.7\% \\
 Output coupling efficiency of PPLN waveguide & & 85.7\% \\
  \hline
  \hline
\end{tabular}
\end{table}

\begin{table}\color{black}
	\caption{Experimental parameters in our system.}
	\label{tables2}
	\begin{ruledtabular}
		\begin{tabular}{c@{}c@{}c@{}}
			
		Parameter & Description & Result \\ \hline
			$R_{rep}$	& Repetition rate of qubit. & 500 MHz  \\	
			$\mu _{\mathrm{SPDC}}$	& Average entangled photon pair per qubit & 0.042  \\
			$\mu _{\mathrm{A}}$	& Average teleported photon  per qubit before BSM & 0.029  \\
			$t_{\mathrm{i}}$	& Transmission probability of teleported photons & 0.147  \\
			$t_{\mathrm{i}}$	& Transmission probability of idler photons & 0.012  \\
			$t_{\mathrm{s}}$	& Transmission probability of signal photons & 0.014  \\
			$\xi _{\mathrm{BSM}}$	& Detection efficiency of the SNSPD for BSM & 0.60  \\
			$\xi _{\mathrm{s}}$	& Detection efficiency of the SNSPD for signal photons & 0.80  \\
			$\zeta$	& Indistinguishability of BSM photons & 0.89 ± 0.02  \\
		\end{tabular}
	\end{ruledtabular}
\end{table}

\begin{table}\color{black}
	\caption{Gains[Hz] for different input states and mean photon number.}
	\label{tables3}
	\begin{ruledtabular}
		\begin{tabular}{c@{}c@{}c@{}c@{}}
			
		Input state & signal & decoy & vacuum \\ \hline
			$|e\rangle$	& 9.92 ± 0.22 & 3.78 ± 0.14 & 0.66 ± 0.06  \\	
			$|l\rangle$	& 10.35 ± 0.23 & 4.01 ± 0.14 & 0.66 ± 0.06  \\
			$|+\rangle$	& 5.76 ± 0.17 & 1.92 ± 0.10 & 0.34 ± 0.04  \\
			$|+i\rangle$	& 6.04 ± 0.17 & 2.01 ± 0.10 & 0.32 ± 0.44 \\
		\end{tabular}
	\end{ruledtabular}
\end{table}

\clearpage
\begin{table}\color{black}
	\caption{Fidelities for different input states and mean photon number.}
	\label{tables4}
	\begin{ruledtabular}
		\begin{tabular}{c@{}c@{}c@{}c@{}c@{}}
			
		Input state & signal & decoy & vacuum & single-photon \\ \hline
			$|e\rangle$	& 92.8 ± 0.6\% & 90.3 ± 1.1\% & 53.4 ± 4.4\% & $\geqslant$ 97.8 ± 1.8\%  \\	
			$|l\rangle$	& 91.9 ± 0.6\% & 89.7±1.1\% & 53.0 ± 4.4\% & $\geqslant$ 96.6 ± 1.7\% \\
			$|+\rangle$	& 76.1 ± 1.3\% & 84.9 ± 1.9\% & 56.5 ± 6.2\% & $\geqslant$ 89.7 ± 3.0\%  \\
			$|+i\rangle$	& 74.1 ± 1.3\% & 81.1 ± 2.0\% & 52.4 ± 6.3\% & $\geqslant$ 84.9 ± 3.0\% \\
		\end{tabular}
	\end{ruledtabular}
	
\end{table}

\begin{table}\color{black}
	\caption{Gains [Hz] of equatorial states for different $\mu _{\mathrm{SPDC}}$.}
	\label{tables5}
	\begin{ruledtabular}
		\begin{tabular}{c@{}c@{}c@{}c@{}}
			
		$\mu _{\mathrm{SPDC}}$ & signal & decoy & vacuum \\ \hline
			0.005	& 0.70 ± 0.03 & 0.21 ± 0.02 & 0.02 ± 0.01  \\	
			0.01	& 1.34 ± 0.04 & 0.46 ± 0.02 & 0.03 ± 0.01  \\
			0.021	& 2.42 ± 0.04 & 0.92 ± 0.02 & 0.07 ± 0.01  \\
			0.042	& 5.76 ± 0.17 & 1.92 ± 0.10 & 0.34 ± 0.04 \\
			0.063	& 8.90 ± 0.21 & 3.08 ± 0.12 & 0.66 ± 0.06 \\
			0.084	& 12.20 ± 0.25 & 4.42 ± 0.15 & 1.22 ± 0.08 \\
		\end{tabular}
	\end{ruledtabular}
\end{table}

\begin{table}\color{black}
	\caption{Fidelities of equatorial states for different $\mu _{\mathrm{SPDC}}$.}
	\label{tables6}
	\begin{ruledtabular}
		\begin{tabular}{c@{}c@{}c@{}c@{}c@{}}
			
		$\mu _{\mathrm{SPDC}}$ & signal & decoy & vacuum & single-photon\\ \hline
			0.005	& 80.2 ± 1.1\% & 84.1 ± 1.8\% & 53.3 ± 15.3\% & 84.4 ± 2.0\% \\	
			0.01    & 77.6 ± 1.1\% & 83.0 ± 1.7\% & 58.3 ± 22.3\% & 84.1 ± 1.8\% \\
			0.021	& 77.8 ± 1.1\% & 82.4 ± 1.6\% & 59.5 ± 8.5\% & 83.9 ± 1.9\%  \\
			0.042	& 77.8 ± 1.0\% & 80.7 ± 1.7\% & 51.4 ± 6.0\% & 84.0 ± 2.3\% \\
			0.063	& 75.5 ± 0.9\% & 79.8 ± 1.5\% & 54.0 ± 4.0\% & 84.2 ± 2.6\% \\
			0.084	& 75.7 ± 0.8\% & 78.7 ± 1.3\% & 54.8 ± 3.2\% & 84.5 ± 2.1\% \\
		\end{tabular}
	\end{ruledtabular}
\end{table}

\clearpage

\end{document}